# Phase Stability of Hexagonal/cubic Boron Nitride Nanocomposites

Abhijit Biswas,[1,*] Rui Xu,[1] Joyce Christiansen-Salameh,[2] Eugene Jeong,[2] Gustavo A. Alvarez,[2] Chenxi Li,[1] Anand B. Puthirath,[1] Bin Gao,[3] Arushi Garg,[4] Tia Gray,[1] Harikishan Kannan,[1] Xiang Zhang,[1] Jacob Elkins,[1] Tymofii S. Pieshkov,[1,5] Robert Vajtai,[1] A. Glen Birdwell,[6] Mahesh R. Neupane,[6] Bradford B. Pate,[7] Tony Ivanov,[6] Elias J. Garratt,[6] Pengcheng Dai,[3] Hanyu Zhu,[1,*] Zhiting Tian,[2,*] and Pulickel M. Ajayan[1,*]

**AFFILIATIONS**

[1]Department of Materials Science and Nanoengineering, Rice University, Houston, TX, 77005, USA

[2]Sibley School of Mechanical and Aerospace Engineering, Cornell University, Ithaca, NY 14853, USA

[3]Department of Physics and Astronomy, Rice University, Houston, TX, 77005, USA

[4]Department of Materials Science and Engineering, Indian Institute of Technology Kanpur, Kanpur, 208016, India

[5]Applied Physics Graduate Program, Smalley-Curl Institute, Rice University, Houston, TX, 77005, USA

[6]DEVCOM Army Research Laboratory, RF Devices and Circuits, Adelphi, Maryland 20783, USA

[7]Naval Research Laboratory, Washington DC 20375, USA

[*]Corresponding Authors: **01abhijit@gmail.com, hanyu.zhu@rice.edu, zhiting@cornell.edu, ajayan@rice.edu**

**ABSTRACT**

Boron nitride (BN) is an exceptional material and among its polymorphs, two-dimensional (2D) hexagonal and three-dimensional (3D) cubic BN (h-BN and c-BN) phases are most common. The phase stability regimes of these BN phases are still under debate and phase transformations of h-BN/c-BN remain a topic of interest. Here, we investigate the phase stability of 2D/3D h-BN/c-BN nanocomposites and show that the co-existence of two phases can lead to strong non-linear optical properties and low thermal conductivity at room temperature. Furthermore, spark-plasma sintering of the nanocomposite shows complete phase transformation to 2D h-BN with improved crystalline quality, where 3D c-BN grain sizes governs the nucleation and growth kinetics. Our demonstration might be insightful in phase engineering of BN polymorphs based nanocomposites with desirable properties for optoelectronics and thermal energy management applications.

## Cover Art

Boron nitride (BN) polymorphs with unprecedented functionalities are important in technological and industrial applications. We synthesized 2D/3D (hexagonal/cubic) h-BN/c-BN nanocomposites that exhibit excellent properties, valuable for the design and engineering of BN-based polymorphs for various technological applications.

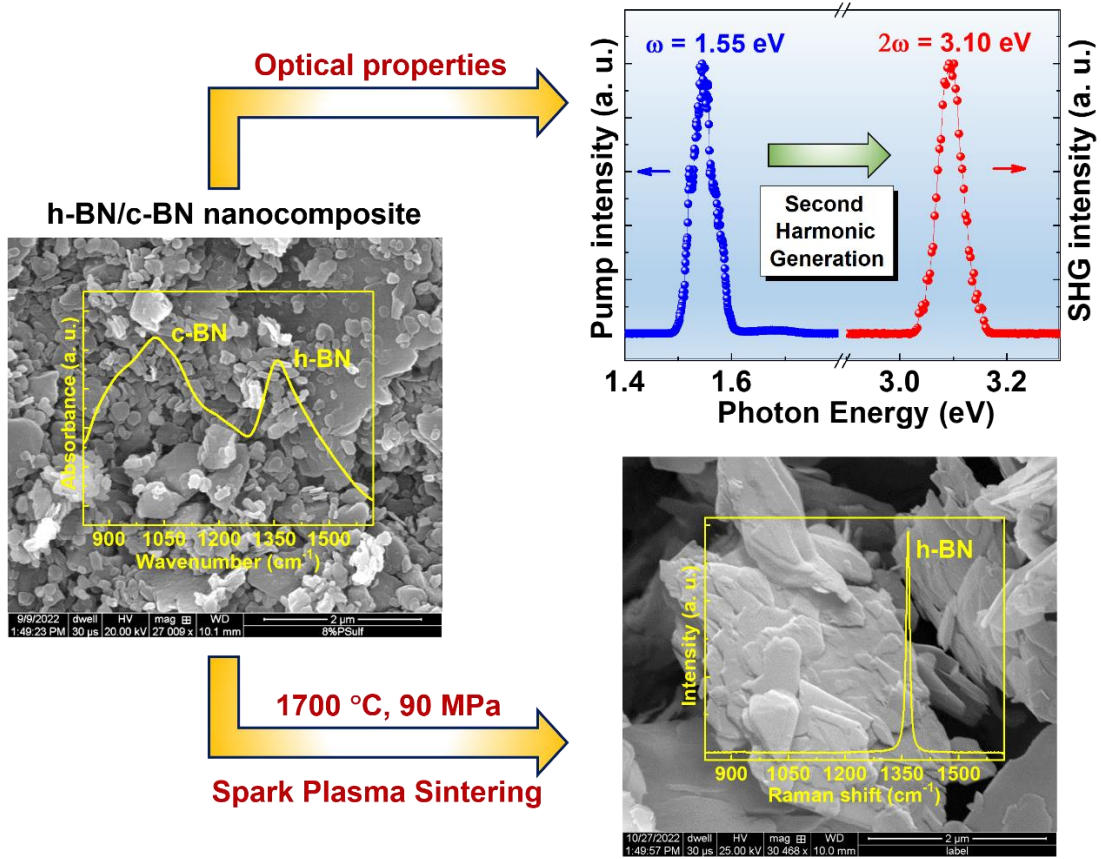

Boron nitride (BN) is one of the most intriguing classes of technologically promising materials due to its unprecedented structural, chemical, thermal, mechanical, optical and electrical properties **[1-6]**. It is unique because of its various structural polymorphs, e.g. hexagonal BN (h-BN), cubic BN (c-BN), rhombohedral BN (r-BN), and wurtzite BN (w-BN) with their consequent functionalities, and nanotechnology/industrial applications **[2-11]**. Between the two common polymorphs, conceivably, two-dimensional (2D) h-BN is regarded as the most stable in ambient conditions **[1]**, whereas three-dimensional (3D) c-BN synthesis requires high pressure and high temperature (HPHT) conditions **[3-5]**. Despite tremendous application potential in modern materials design, synthesis of different polymorphs of BN remains difficult due to the non-trivial thermodynamics and growth kinetics **[12-21]**. Several reports state that at ambient conditions, c-BN is the most stable phase and the h-BN↔c-BN phase transformation occurs within a broad temperature range due to the complex growth kinetics, depending on the grain size and impurity effects **[16-19]**. In contrast, several reports, demonstrate that the conversion of $sp^2$-hybridized h-BN into $sp^3$-hybridized c-BN occurs only at HPHT **[5, 22, 23]**. Therefore, fundamental understanding and exploration of the complex P-T phase diagram of h-BN↔c-BN remains an exciting arena of research for BN polymorphs.

Amongst BN polymorphs, structurally, 2D h-BN is a layered van-der Waals (vdW) material with hexagonal unit-cell having lattice parameters of $a$ = 2.504 Å and $c$ = 6.661 Å **[2]**. Bulk h-BN is centrosymmetric (space group: P63/mmc) and its most stable, lowest energetic orientation is along [002]. It is a chemically inert electrically insulating material with an ultrawide-bandgap of ~5.9 eV, and exhibits thickness-dependent electronic properties **[2]**. Its weak inter-layer bonding makes it a soft lubricant material and thus promising as a high-temperature corrosion resistant and antioxidative coating for various industrial applications **[24]**. Due to its low dielectric constant and high dielectric breakdown strength, h-BN is also very useful for 2D-electronics as a capping or dielectric layer **[25]**. In addition, h-BN has been used for gas sensing (e.g. ammonia, ethanol) and related energy storage applications **[26, 27]**. On the other hand, 3D c-BN forms a zinc blende structure with cubic unit-cell (analogous to diamond) having lattice parameters of $a$ = 3.62 Å (space group: $\bar{F}$43m). Its most stable, lowest energetic facet is along [111], which is polar **[5]**. It has a much wider bandgap of ~6.3 eV **[5, 28]**. c-BN is also technologically important as an excellent abrasive and machining tool exceeded only by diamond), because of its exceptionally high Vickers hardness and thermal conductivity **[5, 29]**. Furthermore, the chemical inertness of c-BN makes it suitable for thermocouple protection sheaths, crucibles, and protective linings for reaction vessels **[5, 30]**. Comparatively, the density of h-BN is ~2.1 g/cm$^3$, whereas that for c-BN is ~3.45 g/cm$^3$ **[5]**. Therefore, considering the diverse functionalities and tremendous application worthiness, combining 2D h-BN and 3D c-BN and making a 2D/3D h-BN/c-BN nanocomposite would generate novel materials with tailored diverse properties **[31]**, which might provide pivotal insights into the material design and engineering of BN polymorphs.

Here, we have synthesized the 2D/3D nanocomposite by mixing the h-BN and c-BN powders and performed structural, optical, and thermal conductivity characterizations. Comprehensive structural characterizations confirm the presence of both the h-BN and c-BN phases. Optically, the nanocomposite is second harmonic generation (SHG) active and the power scaling of SHG follows quadratic dependence. The room temperature thermal conductivity of the nanocomposite is found to be ~1.9 W/(mK). We performed the spark plasma sintering (SPS) of the nanocomposite and observed that 2D/3D h-BN/c-BN nanocomposite transforms to 2D h-BN. These observations are important for exploring the h-BN↔c-BN phase transformation as well as for the design and engineering of BN polymorph-based 2D/3D nanocomposites for innumerable applications.

First, we characterized the commercially available h-BN and c-BN powders by using several techniques. **Figure 1** shows X-ray diffraction (XRD), X-ray photoelectron spectroscopy (XPS), field emission scanning electron microscope (FESEM), Raman spectroscopy, and Fourier-transform infrared spectroscopy (FTIR) of pristine 2D h-BN (left panel) and 3D c-BN (right panel). As shown, in XRD (**Figure 1a**) both show the characteristic diffraction peaks, with the most intense ones for (002) of h-BN and (111) of c-BN, the most stable structure of respective phases, signifying that most of the grains are oriented along these directions. XPS shows the characteristic B-N peaks (at B1s and N1s core) for both cases, however, an additional π-Plasmon peak appears for h-BN (at ~9 eV from the B-N peak), but not for the case of c-BN (**Figure 1b**) **[32-35]**. We also performed the XPS valence band spectroscopy (VBS) of h-BN and c-BN (supplementary **Figure S1**). The VBS of the c-BN exhibits a similar shape as observed for h-BN, with a moderate shift of valence band maxima (VBM) towards lower the binding energy from the Fermi level ($E_F$) **[36, 37]**. FESEM shows 2D sheet-like features for h-BN (sheets of few μm), whereas 3D-islands for c-BN (grain size <1 μm) (**Figure 1c**). Raman spectra shows the $E_{2g}$ phonon mode (~1366 cm$^{-1}$) of

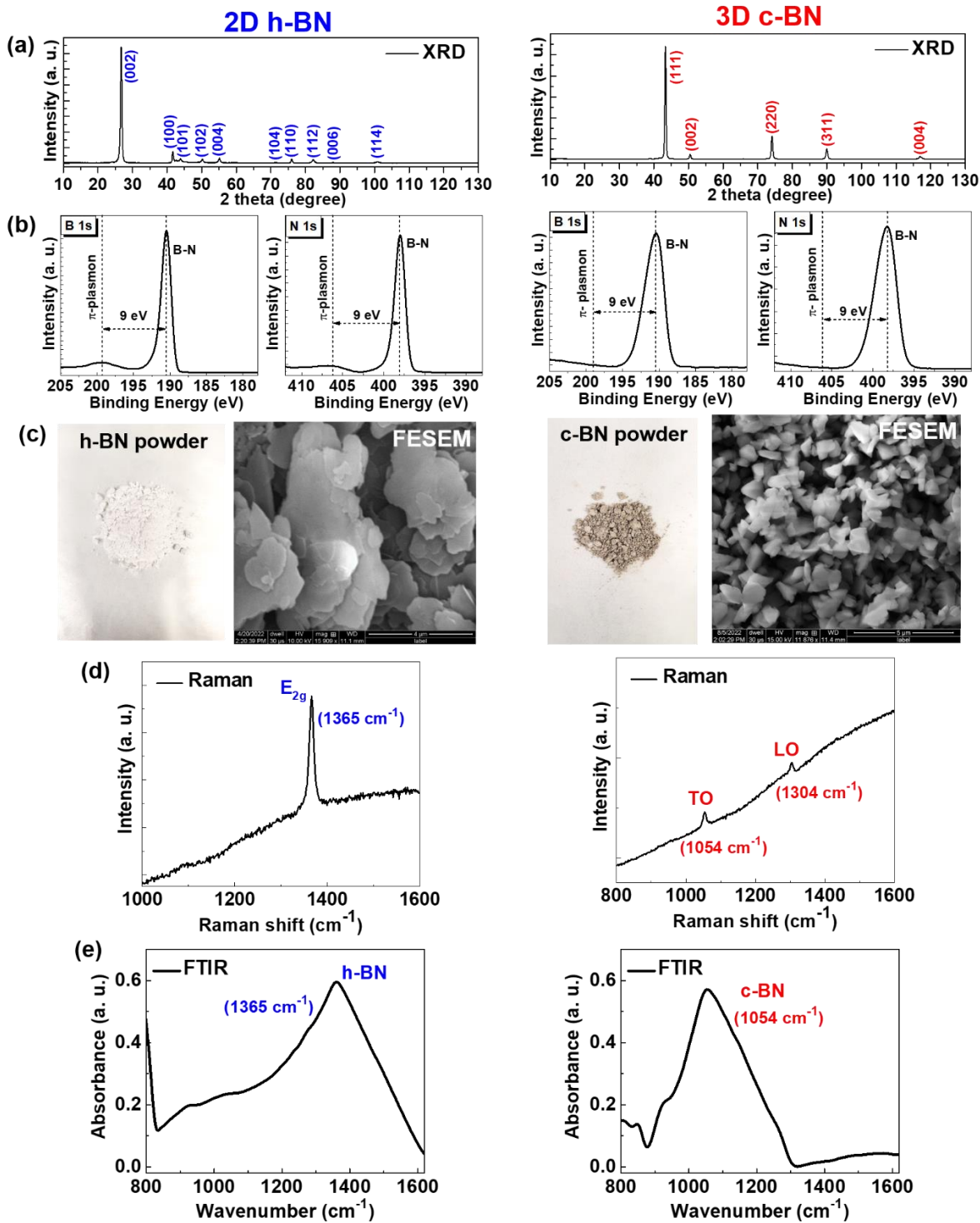

**Figure 1: Characterizations of bulk 2D h-BN and 3D c-BN powder.** (a)-(e) X-ray diffraction pattern, X-ray photoelectron spectroscopy, Field-emission scanning electron microscopy, Raman spectra and Fourier-transform infrared spectroscopy characterizations of the respective 2D h-BN (left panel) and 3D c-BN phase (right panel).

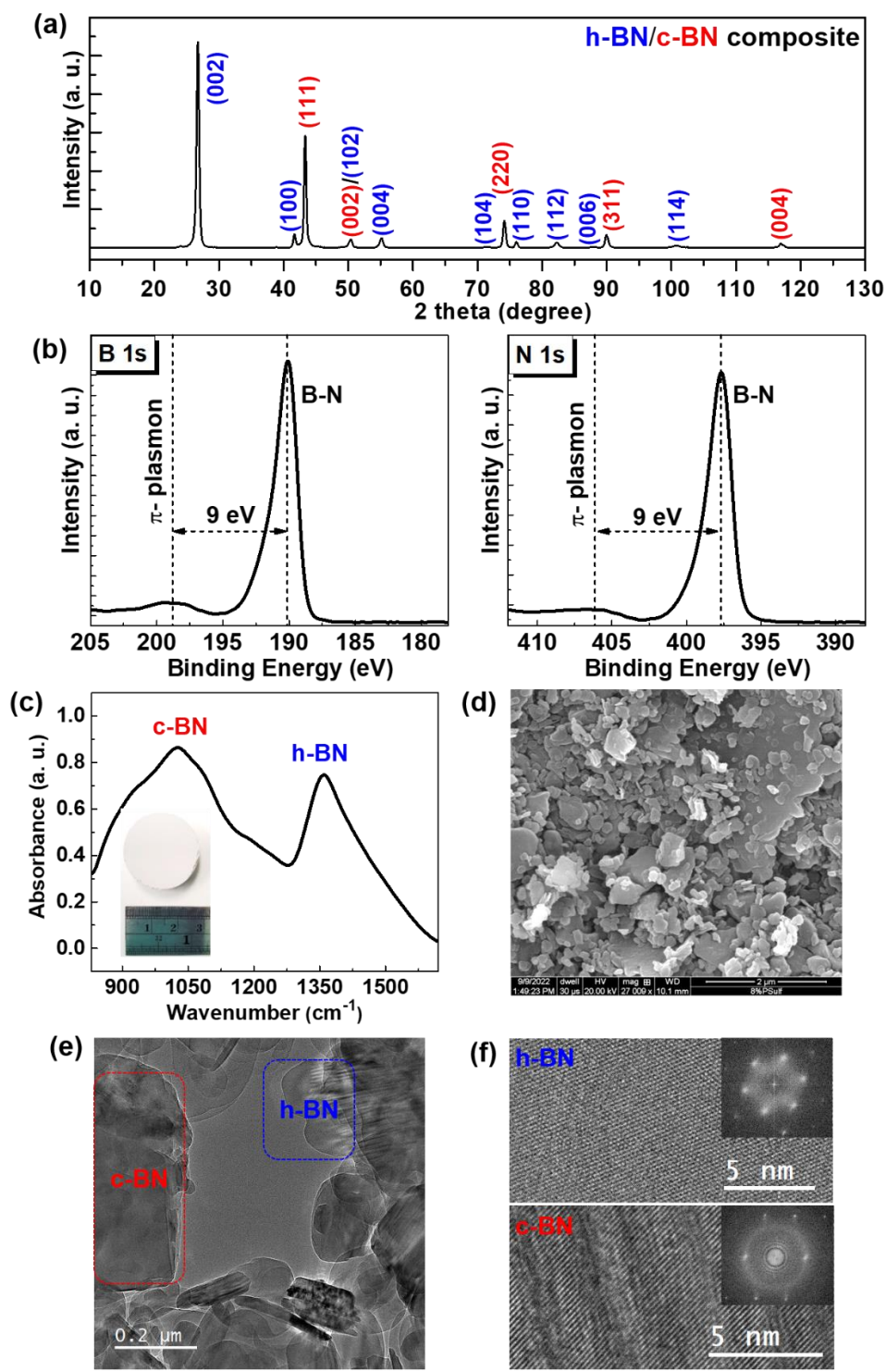

**Figure 2: Characterizations of the h-BN/c-BN nanocomposite.** (a)-(f) XRD, XPS, FTIR, FESEM, and HRTEM shows the co-existence of h-BN (blue color) and c-BN (red color) phases. Inset of (c) showing the sintered one-inch white h-BN/c-BN pellet. (f) HRTEM lattice fringes and corresponding diffraction patterns (inset) further confirm the co-existence of both of the phases.

sp$^2$ bonded h-BN, whereas transverse optical (TO ~1054 cm$^{-1}$) and longitudinal optical (LO ~1304 cm$^{-1}$) vibrational modes for c-BN (**Figure 1d**). These findings are further supported by the FTIR spectra observed for the respective pristine 2D h-BN and 3D c-BN (**Figure 1e**) **[28, 32, 33, 38]**.

Subsequently, we made the 2D h-BN and 3D c-BN based nanocomposite (2D/3D h-BN/c-BN) with 1:1 molar ratio by using the standard solid-state reaction method and sintered it at 1000 °C for 12 hrs in a vacuum-sealed quartz tube. XRD shows the diffraction peaks from both h-BN (indexed in blue) and c-BN (indexed in red) (**Figure 2a**). XPS elemental scans at both B1s and N1s core shows the characteristic B-N peaks (**Figure 2b**). FTIR spectra show the presence of both h-BN and c-BN (**Figure 2c**). From FTIR absorbance intensity, we estimated the c-BN and h-BN content in the nanocomposite of ~0.52:0.48 **[28]**. The image of an as-sintered one-inch h-BN/c-BN pellet is shown (inset of **Figure 2c**). In addition, the FESEM image shows the coexistence of both sheet-like (h-BN) and island-like (c-BN) grains (**Figure 2d**). We examined the structure of the composite by putting it onto the Cu-grid (see **method section**) and measuring it with high-resolution transmission electron microscopy (HRTEM). We observed the presence of both h-BN and c-BN (**Figures 2e, 2f** and supplementary **Figure S2**) (with their lattice fringes and corresponding ((0002) h-BN and (111) c-BN)) diffraction patterns **[1, 39]**. These characterizations indeed confirm the formation of 2D/3D h-BN/c-BN nanocomposite.

While bulk h-BN is centrosymmetric, c-BN breaks the inversion symmetry **[39-42],** thus finite second-order nonlinear optical effect can be expected for the nanocomposite. Hence, we performed the optical nonlinear second harmonic generation (SHG) of the nanocomposite (more details in the **method section**). Under 800 nm (~1.55 eV) pumping, we observed a giant SHG signal with exact double frequency (~3.10 eV) from the composite surface (**Figure 3a**). By inserting an optical filter to eliminate the pumping laser, the SHG emission can even be directly seen on a charge-coupled device (CCD) camera without any sample damage with a pumping laser intensity of ~7 GW/cm$^2$. Power dependence measurement also reveals a nearly quadratic relationship between SHG and incident laser power. Fitting the power dependence with $I_{SHG} = AP_{pump}^k$ **[43]** gives the exponent k = 2.042±0.015 for the h-BN/c-BN nanocomposite (**Figure 3b**).

We compared the SHG yield with the pristine c-BN or h-BN powder, which also showed SHG response, however it is one-order of magnitude lower than for the h-BN/c-BN nanocomposite (**Figure 3b**). The SHG enhancement from the nanocomposite might be attributed to the high-temperature annealing (at 1000 °C) which possibly helps in improving the crystallinity, and combining the effect of the two phases; thus effectively improving the SHG efficiency. To quantify the optical nonlinearity of the nanocomposite, we first calibrated the SHG response using a monolayer 2H-WS$_2$ sample as a reference (a highly non-centrosymmetric material with high second-order nonlinear optical susceptibility) [43]. Here, single crystal WS$_2$ is mechanically exfoliated onto SiO$_2$/Si substrate, and the second-order nonlinear optical susceptibility $|\chi^{(2)}|$ is calculated to be ~1.2 nm/V, showing good agreement with the previous experimental result [43, 44]. SHG yield of our h-BN/c-BN nanocomposite is comparable to that of the monolayer 2H-WS$_2$ (**Figure 3b**), with an SHG intensity ratio P$_{c-BN/h-BN}$:P$_{WS2}$ ~1:2. The effective interaction length of forward-propagating SHG inside the sample is $d = \frac{1}{\Delta k} = \frac{\lambda_{SHG}}{2\pi(n_{SHG}-n_{pump})}$ ~ 300 nm, where $\Delta k$ is the phase mismatch due to the difference in refractive indices (RI) at the pump ($n_{pump}$) and SHG ($n_{SHG}$) frequencies.[45, 46] Thus, we estimate that the lower bound of the effective second-order nonlinear susceptibility $|\chi^{(2)}|$ in the composite is about of ~ 3 pm/V. This is in the same order as the values in typical nonlinear optical crystals and possibly makes the nanocomposite useful for non-linear optoelectronic applications.

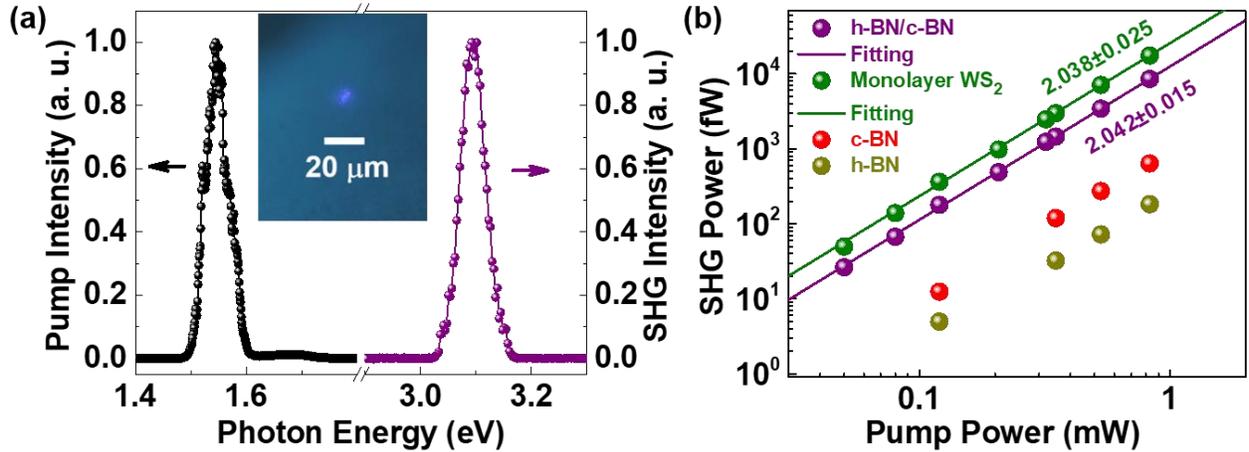

**Figure 3: Optical second harmonic generation of h-BN/c-BN nanocomposite.** (a) Two lower energy photons are up-converted to exactly twice the incident frequency of an excitation laser. The SHG emission can be directly seen from the CCD camera (inset) (b) Pump power dependence of SHG power intensity shows the quadratic dependence (purple color). For comparison, we have shown power dependence for monolayer 2H-WS$_2$, pristine c-BN, and h-BN.

The thermal conductivity ($k$) of BN is also important for various applications as it can act as a heat sink or thermal isolation material. Therefore, we obtained the room temperature ($k$) of the composite by measuring the thermal diffusivity ($\alpha$) with the laser flash method, the specific heat capacity ($C_p$) with differential scanning calorimetry (**Figure 4a**) (see **method section**), and the density ($\rho$) with Archimedes method, and applying the relation $k = C_p * \rho * \alpha$. The laser flash method data (temperature rise vs. time) was fitted with the Dusza combined model (**Figure 4b**) [**47**]. We obtained $\alpha$ = 0.0135±0.00056 cm²/s, $C_p$ = 0.903±0.0135 J/(g°C), and $\rho$ = 1.57±0.007 g/cm³, yielding $k$ = 1.91±0.08 W/(mK). In literature, BN ceramics composed solely of h-BN have higher thermal conductivity [**48, 49**]. Arguably, this is the first demonstration of the thermal conductivity measurement of BN polymorph-based nanocomposite. We attribute this lower $k$ of the nanocomposite to the low relative density of ~57% (calculated as the percentage of the theoretical density of the h-BN/c-BN nanocomposite), and to the thermal resistance between the 2D/3D h-BN/c-BN grains, the effect of which is amplified by the increased concentration of grain boundaries due to presence of small c-BN grains. Nevertheless, low thermal conductivity could be useful for the generation and efficient management of thermal energy acting as thermal isolation, and might be helpful for achieving a high thermoelectric figure of merit in designed thermoelectric materials based on nanocomposites.

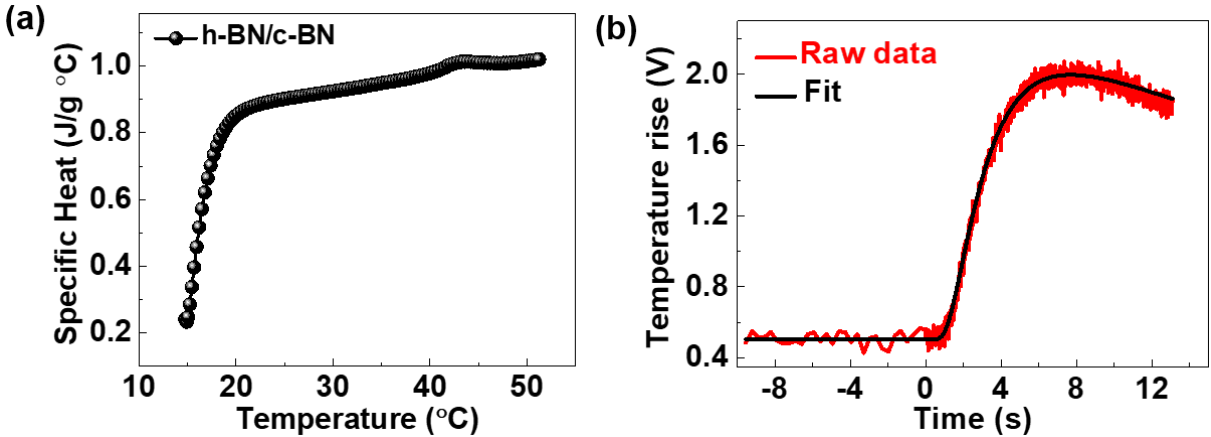

**Figure 4: Specific heat capacity and laser flash method data of h-BN/c-BN nanocomposite.** (a) Temperature-dependent specific heat capacity of the h-BN/c-BN nanocomposite obtained by the differential scanning calorimetry. (b) Dusza combined model fitting of the laser flash method temperature rise vs. time data from which the room temperature thermal diffusivity is obtained.

Furthermore, we performed spark plasma sintering (SPS), a pressure-assisted sintering process to densify the ceramics (see **method section**). The comparative structural characterizations of the composite, before (left panel) and after (right panel) the SPS are shown (**Figure 5**). Structurally, XRD, FESEM, Raman, FTIR, and HRTEM show the presence of only h-BN, after the SPS. XRD shows the diffraction peaks all correspond to the h-BN (**Figure 5a**). Remarkably, the quality of h-BN is much improved as the (0002) h-BN peak becomes much narrower with FWHM of ~0.503° (before the SPS) and ~0.303° (after the SPS) (supplementary **Figure S3**). After the SPS, the VBM shifts away from the Fermi-level (w.r.t. the as-synthesized h-BN/c-BN), and becomes very similar to the pristine h-BN (supplementary **Figure S1**). FESEM shows the sheet-like feature, evident of h-BN (**Figure 5b**). Raman spectra only show the h-BN $E_{2g}$ peak at ~1367 cm$^{-1}$ (**Figure 5c**). The full-width at half maximum (FWHM) of the Raman peak was found to be ~13 cm$^{-1}$, further confirms excellent crystalline quality of h-BN **[28, 32]**. FTIR also shows only the h-BN-related peak (**Figure 5d**). Moreover, HRTEM shows the presence of sheet-like features, corresponding to 2D h-BN (**Figure 5e** and supplementary **Figure S3**). All these observations confirms that after the SPS, 2D/3D h-BN/c-BN nanocomposite fully transforms into 2D h-BN phase.

Considering the BN P-T phase diagram, this transformation is quite unusual as reports show inconsistencies about the c-BN↔h-BN phase transformation. Various reports had shown that at ambient conditions, c-BN is the most stable phase and it converts to h-BN at higher temperatures and pressure **[14-19, 50, 51]**. However, Solozhenko *et al.*, calculated that c-BN→h-BN conversion occurs ~1000-1800 °C, as vapor pressure increases with the temperature due to the endothermic reaction **[14]**. Additionally, Wolfrum *et al.* had shown that SPS done under low pressure could convert 3D c-BN into 2D h-BN due to the presence of possible boron oxide ($B_2O_3$) impurities (acting as a catalyst) which can play a role in phase transformations **[52]**. Sachdev *et al.* found that c-BN grain size and the presence of $B_2O_3$ influence the phase conversion from c-BN to h-BN, with the conversion temperature of ~900 °C for smaller c-BN grains (~1.5 μm) and ~1500 °C for larger c-BN grains (~600 μm), which influences the activation energies and kinetic factors **[53]**. Cahill *et al.* investigated the c-BN particle size-dependent transformation under the helium atmosphere and revealed that the growth of h-BN increases with both time and temperature (~1560-1660 °C) **[54]**. In view of the impurity phase, before and after the SPS, we did not observe any $B_2O_3$ peak (supplementary **Figure S4**).

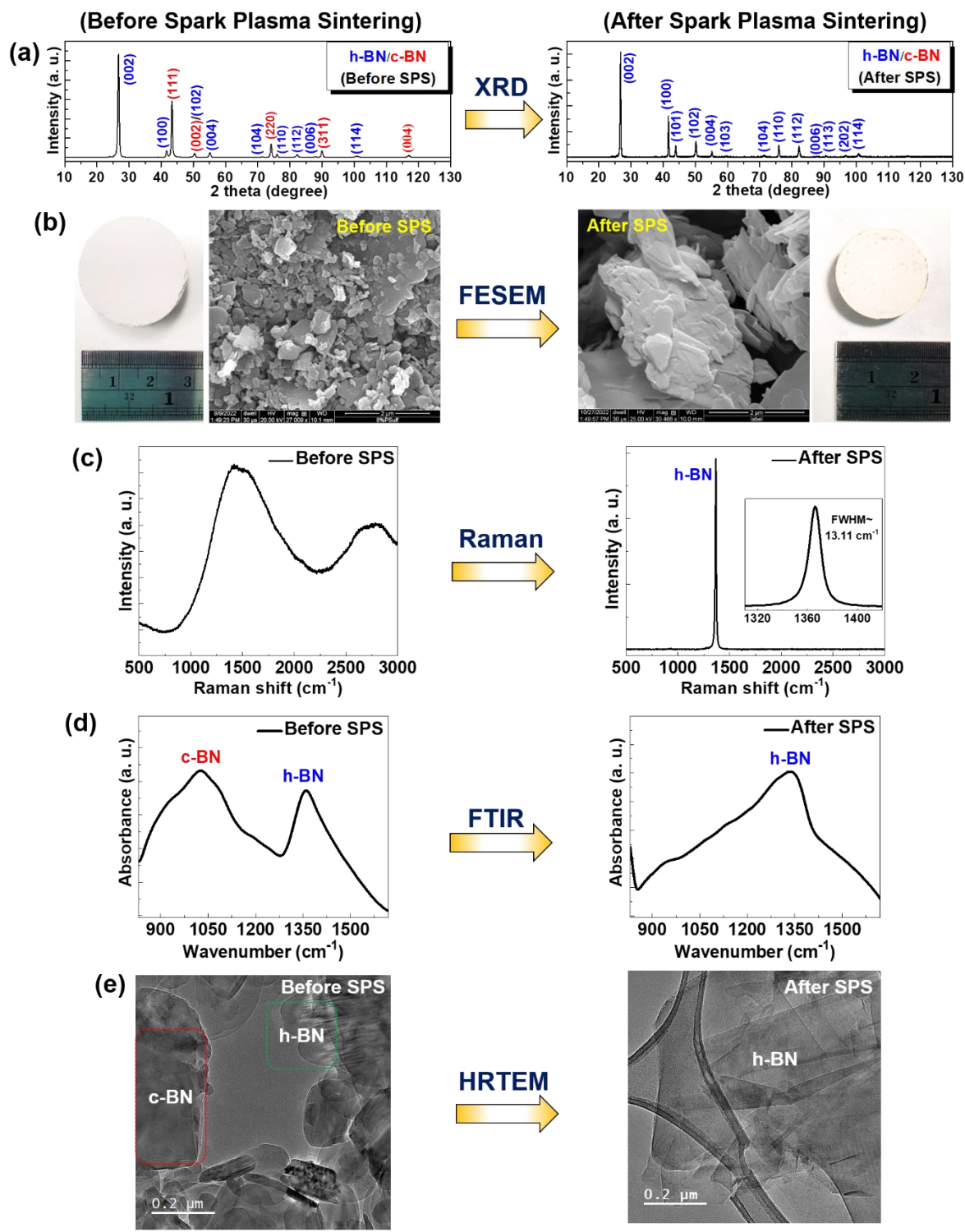

**Figure 5: Characterizations of h-BN/c-BN nanocomposite after the spark plasma sintering.** (a)-(f) XRD, FESEM, Raman, FTIR, and HRTEM shows the complete phase transformation of 2D/3D h-BN/c-BN nanocomposite to h-BN, after the spark plasma sintering at 1700 °C and 90 MPa pressure. The individual disk is shown (the scale bar is in cm scale).

Interestingly, the FWHM of the (0002) h-BN peak reduces after the SPS, indicating the improved crystalline quality of h-BN. In terms of grain size, in our case, c-BN grain sizes are <1 µm (**Figure 1c**). Thus, considering the lower conversion temperature for smaller size c-BN grains **[51, 53]**, the conversion temperature would probably below the temperature used for the SPS (~1700 °C). To confirm this, we also performed the SPS at room temperature and 1000 °C (at 90 MPa), however, the nanocomposite does not show any significant structural changes, and except a minor difference in absorbance intensity for the respective BN phases (supplementary **Figures S5** and **S6**). The density of BN phases is also a crucial factor for the phase transformation **[55]**, as according to the Ostwald-Volmer rule, a lesser dense phase (h-BN density is lower than the c-BN) should nucleate first and the growth always precedes by nucleation **[53]**. For nanocrystals, the surface energy becomes more dominant in the total free energy **[56]**. During the phase transformation, a system proceeds from higher to lower energies, overcoming the energy barrier off the transformation. In that scenario, the phase with lower surface energy (~35-37 mJ/m$^2$ for (002) h-BN, whereas ~3.1 J/m$^2$ for (111) c-BN) **[57, 58]** becomes more energetically favorable. Therefore, the transformation of h-BN/c-BN nanocomposite to h-BN by SPS treatment is attributed with the smaller size c-BN related nucleation kinetics.

In summary, we have synthesized the nanocomposite of 2D/3D h-BN/c-BN and confirmed the co-existence of both phases by performing extensive structural characterizations. Optically, the nanocomposite shows strong nonlinear optical second harmonic generation response. We obtained low thermal conductivity of the nanocomposite at room temperature. Furthermore, spark plasma sintering of the nanocomposite shows its complete transformation to h-BN with an improved crystalline quality, attributed to the nucleation and activation kinetics of c-BN grain sizes. Our findings provide fundamental insights as well as application worthiness of the nanocomposite, which might be useful as a guideline in designing novel nanocomposite based on BN polymorphs, giving rise to the phase-engineered diverse properties for relevant applications, in addition to exploring the non-trivial BN phase diagram.

# Experiential Methods

*Synthesis of h-BN/c-BN nanocomposites (Solid-state reaction and spark plasma sintering)*

We used commercially available high-purity (99.9% metal basis) h-BN and c-BN superabrasive mircopowder powders, purchased from MSE suppliers, USA. The c-BN particle sizes are <1 μm. The as-purchased powders are mixed in a molar ratio of (1:1), grounded in an agate mortar and pestle for ~30 min by adding a few drops of polyvinyl alcohol (PVA) as a binder. The homogeneously mixed powder was then high pressed (4 Ton Load) to make a compact one-inch diameter pellet. The as-made pellet was then sealed inside a quartz tube (in vacuum), and sintered at 1000 °C for 12 hrs in a box furnace. The ramping up and down rate was kept at ~100 °C/hr. The Spark plasma sintering (SPS) of in-house sintered pellets was carried out on an SPS 25-10 machine (Thermal Technology LLC, California USA) at a constant Uniaxial pressing pressure of 90 MPa and heating rate of 50 °C/min (at SPS facility in Texas A & M University, USA). The maximum temperature did not exceed 1700 °C. Sintering was carried out according to the following scheme: 3.5 g of powder was placed in a graphite mold (diameter of 20 mm) and then placed in the sintering chamber under an initial pressure of 5 MPa. It was held at ~ $2\times10^{-5}$ Torr for ~30 min, and then the powder was sintered for one hour under atmospheric pressure of UHP (~99.999%) Argon gas medium. The temperature of the SPS process was controlled by an optical pyrometer Raytek D-13127 (Berlin, Germany). After the sintering, pressure was released slowly at ~5 MPa/min, while the temperature was ramped down at ~100 °C/min.

*Spectroscopic, Chemical, and microscopic characterizations (XRD, XPS, VBS, FESEM, FTIR, Raman, and HRTEM)*

X-ray diffraction (XRD) patterns were recorded by using Rigaku SmartLab thin film X-ray diffractometer (Tokyo, Japan), at 40 kV and 40 mA, by using a monochromatic Cu K$_\alpha$ radiation source (λ= 1.5406 Å) and at the scanning rate of 1°/min. X-ray photoelectron spectroscopy (XPS) was performed by using PHI Quantera SXM scanning X-ray microprobe with a 1486.6 eV monochromatic Al K$_\alpha$ X-ray source. High-resolution core-level elemental B1s and N1s scans were recorded at 26 eV pass energy. The XPS-valence band spectra (XPS-VBS) were acquired by using the pass energy of 69 eV. FTIR was obtained by using the Nicolet 380 FTIR spectrometer, using a single-crystal diamond window. Renishaw inVia confocal microscope was used for the Raman spectroscopy measurements by using a 532 nm laser as the excitation source. The surface topography was obtained by field emission scanning electron microscope (FESEM) (FEI Quanta 400 ESEM FEG). For FESEM, we sputtered ~10 nm gold layer on the insulting BN surface to avoid the charging effect. For the HRTEM, powders from the nanocomposite were dispersed into the Isopropyl Alcohol (IPA) solution and sonicated in an ultrasonic bath for 30 min. Then we dipped the Cu-grid in the solution for a few sec, took it out, and dried it for 30 min. The Cu-grid was then mounted into the TEM chamber and images were recorded using Titan Themis operating at 300 kV.

*Optical Second harmonic generation*

The second harmonic generation was performed with a reflection geometry using the home-built setup. A MaiTai (Spectra-Physics, USA) laser with 84 MHz repetition rate was used to give near-infrared pumping with a wavelength of 800 nm and 40 fs pulse duration after optical compression. The laser is directed to a microscope with a scanning stage, which allows for spatially resolved SHG imaging, and then focused, by a 20× objective with a numerical aperture of 0.45 to a spot size with a diameter of ~9 μm. The average laser power was kept at ~8 mW at the sample location. The reflected signal is filtered by a 785 nm short-pass filter and a 400 nm band-pass filter to eliminate the reflected pump beam. The signal is finally detected using a single-pixel photon counter (C11202-100, Hamamatsu, Japan).

*Thermal conductivity measurement*

The thermal diffusivity of a one-inch diameter, ~5.35 mm thick h-BN/c-BN nanocomposite pellet was measured with laser flash method by using a Linseis XFA 500 Xenon Flash Thermal Conductivity Analyzer. A thin layer of graphite spray coating was applied to the surfaces of the pellet to promote laser absorbance. The measurement was conducted at room temperature with a 10 J laser pulse. The density of the pellet was determined by the Archimedes method. The specific heat capacity of a 39.9 mg portion of the h-BN/c-BN composite was measured using a TA Q200 Differential Scanning Calorimeter. The sample was placed in an aluminum pan, and the specific heat capacity was measured over the temperature range of 15- 50 °C at a heating rate of 10 °C/min.

## ACKNOWLEDGMENTS

This work was sponsored partly by the Army Research Office and was accomplished under Cooperative Agreement Number W911NF-19-2-0269. The views and conclusions contained in this document are those of the authors and should not be interpreted as representing the official policies, either expressed or implied, of the Army Research Office or the U.S. Government. The U.S. Government is authorized to reproduce and distribute reprints for Government purposes notwithstanding any copyright notation herein. R.X. and H.Z. are supported by the U.S. National Science Foundation (NSF) under Award No. DMR 2005096. This work was partly sponsored by the Department of the Navy, Office of Naval Research under ONR award number N00014-22-1-2357. The authors thank Ithaca College for providing access to a differential scanning calorimeter. We would like to thank Dr. Atin Pramanik for providing the PVA binder solution. Authors would also like to acknowledge to SPS facility at Texas A&M University, TX, USA.

## AUTHOR DECLARATIONS

### Conflict of Interest

The authors have no conflicts to disclose.

### Author Contributions

A. B., R. V., and P. M. A. conceptualized the study. A.B., C. L., A. G., T. G., H. K., X. Z., T. P., and J. E. synthesized and characterized the materials. A. B. P. performed the electron microscopy. R. X. and H. Z. carried out the second harmonic optical measurement. J. C., E. J., G. A., and Z. T. measured thermal conductivity. A. G. B., M. R. N., E. J. G. and T. I. commented on the manuscript. All the authors discussed the results and contributed to the manuscript preparation.

## DATA AVAILABILITY

The data that support the findings of this study are available from the corresponding author upon reasonable request.

# Supplementary Materials

# Phase Stability of Hexagonal/cubic Boron Nitride Nanocomposites

Abhijit Biswas,[1,*] Rui Xu,[1] Joyce Christiansen-Salameh,[2] Eugene Jeong,[2] Gustavo A. Alvarez,[2] Chenxi Li,[1] Anand B. Puthirath,[1] Bin Gao,[3] Arushi Garg,[4] Tia Gray,[1] Harikishan Kannan,[1] Xiang Zhang,[1] Jacob Elkins,[1] Tymofii S. Pieshkov,[1,5] Robert Vajtai,[1] A. Glen Birdwell,[6] Mahesh R. Neupane,[6] Bradford B. Pate,[7] Tony Ivanov,[6] Elias J. Garratt,[6] Pengcheng Dai,[3] Hanyu Zhu,[1,*] Zhiting Tian,[2,*] and Pulickel M. Ajayan[1,*]

## AFFILIATIONS

[1]Department of Materials Science and Nanoengineering, Rice University, Houston, TX, 77005, USA

[2]Sibley School of Mechanical and Aerospace Engineering, Cornell University, Ithaca, NY 14853, USA

[3]Department of Physics and Astronomy, Rice University, Houston, TX, 77005, USA

[4]Department of Materials Science and Engineering, Indian Institute of Technology Kanpur, Kanpur, 208016, India

[5]Applied Physics Graduate Program, Smalley-Curl Institute, Rice University, Houston, TX, 77005, USA

[6]DEVCOM Army Research Laboratory, RF Devices and Circuits, Adelphi, Maryland 20783, USA

[7]Naval Research Laboratory, Washington DC 20375, USA

[*]Corresponding Authors: **01abhijit@gmail.com, hanyu.zhu@rice.edu, zhiting@cornell.edu, ajayan@rice.edu**

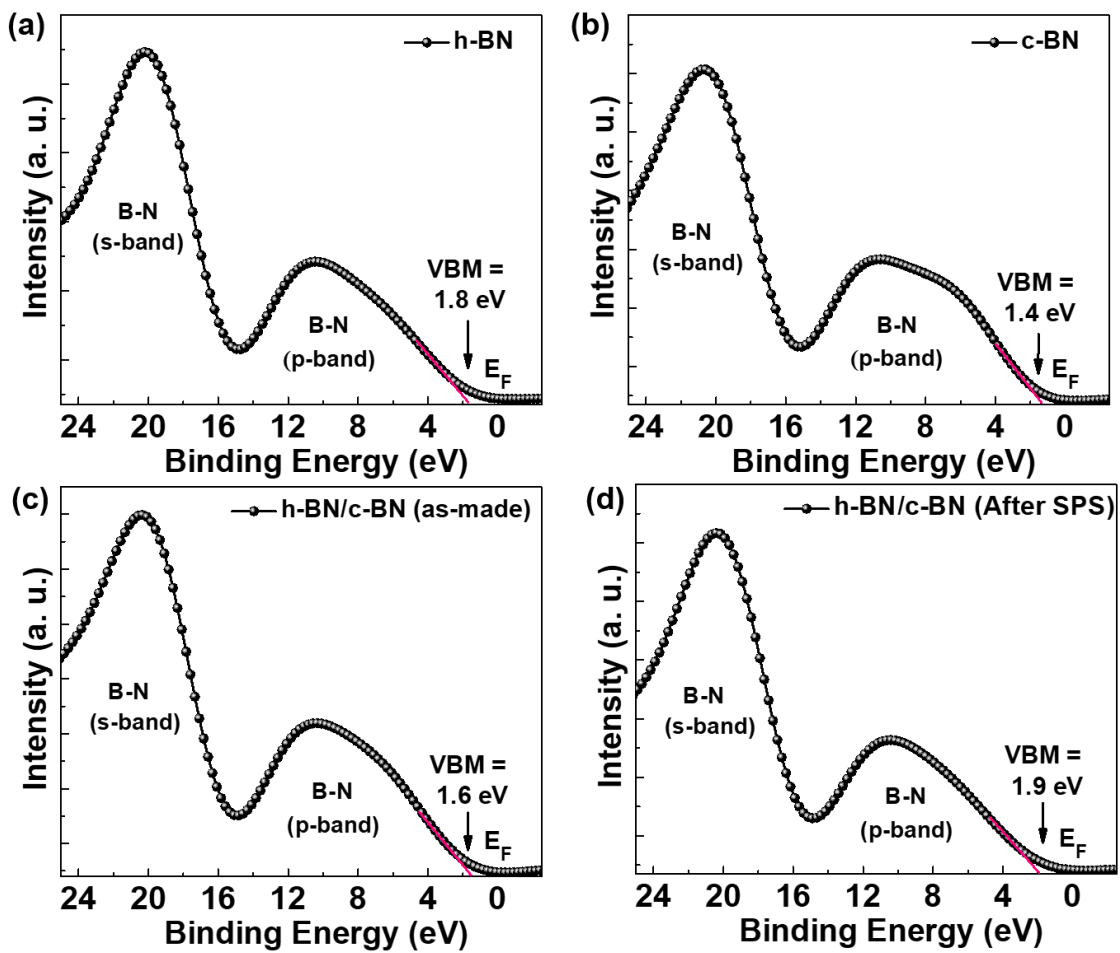

**Figure S1:** (a)-(d) Valence-band spectrum (VBS) of the h-BN, c-BN, h-BN/c-BN nanocomposite, and the nanocomposite after doing the SPS at higher temperature and pressure. The valence band maxima (VBM) positions are shown for the respective sample.

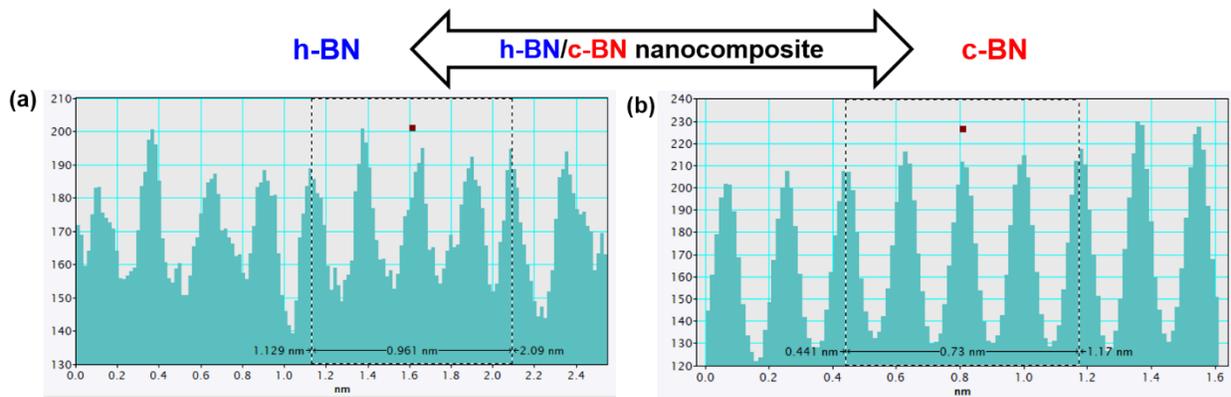

**Figure S2:** (a), (b) Inter-planar distance (*d*) of h-BN and c-BN in the h-BN/c-BN nanocomposite. The average *d* spacing's are ~0.24 nm (for h-BN) and ~0.182 nm (for c-BN), close to the in-plane lattice parameter of h-BN and the B-B bond length of c-BN (111) **[2, 39]**.

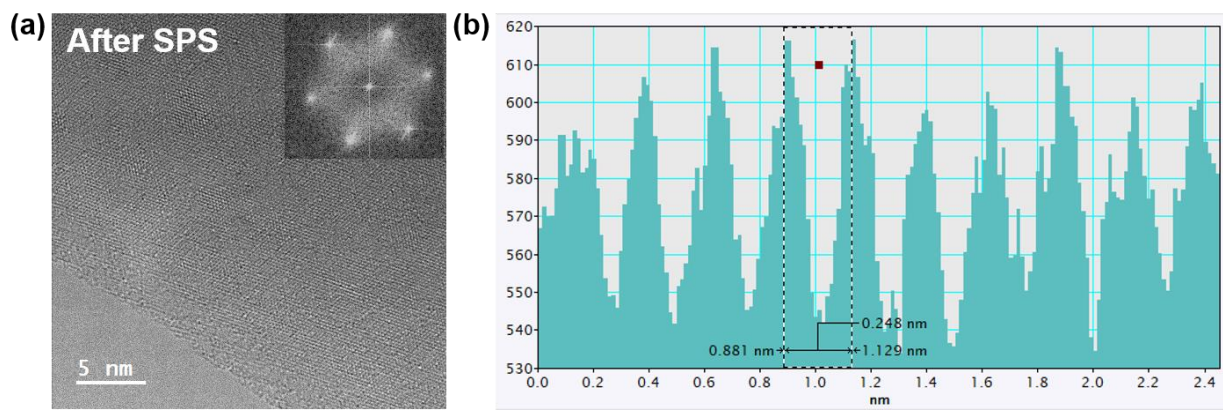

**Figure S3:** (a), (b) Lattice fringes and the corresponding diffraction patterns (inset) of the spark-plasma sintered h-BN/c-BN nanocomposite showing the *d* spacing of ~0.248 nm, corresponding to the in-plane lattice spacing of h-BN.

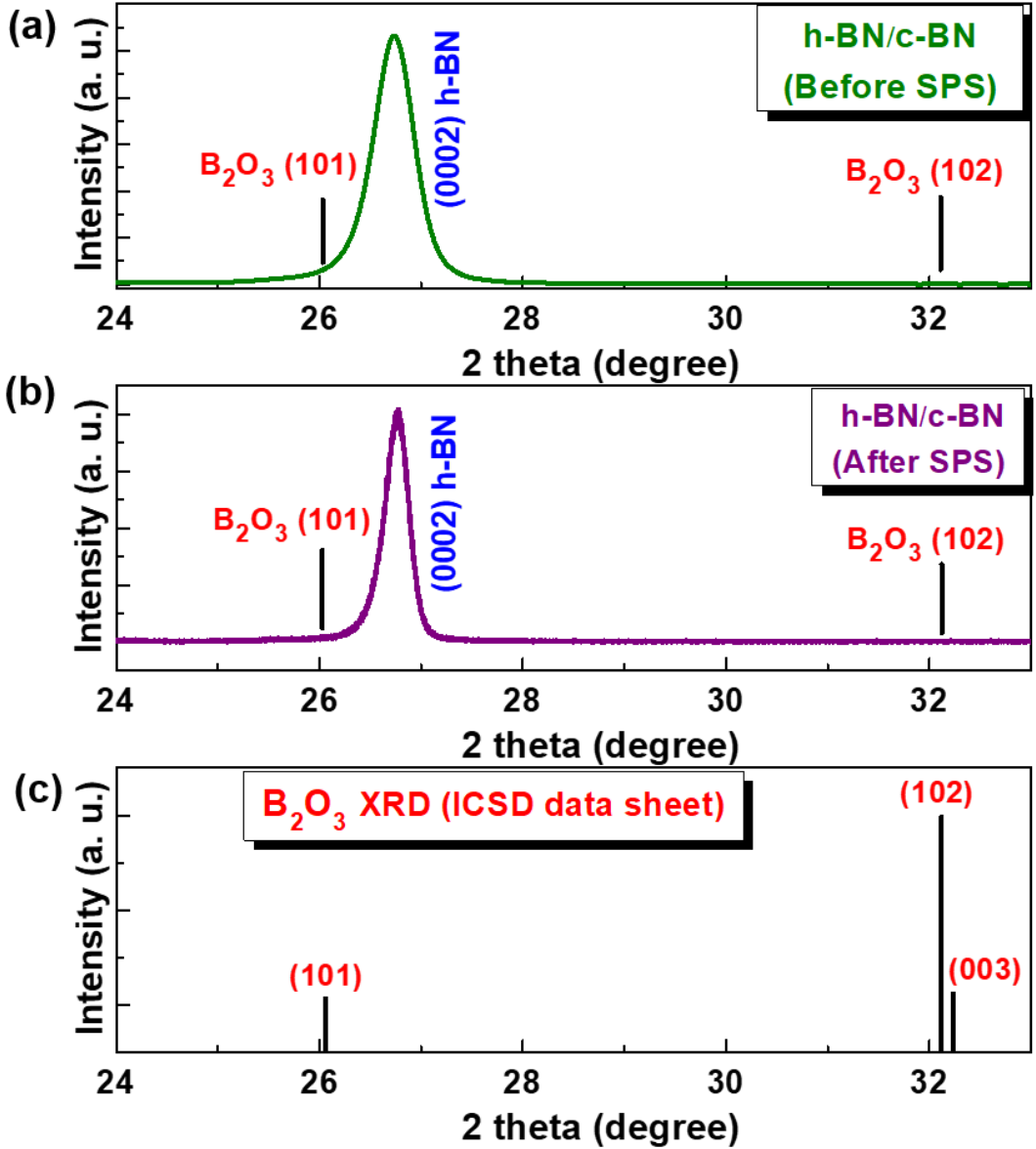

**Figure S4:** (a), (b) XRD to search for the boron oxide ($B_2O_3$, expected at red lines) related impurity peaks, before and after the spark plasma sintering (SPS) of the h-BN/c-BN nanocomposite. (c) XRD of $B_2O_3$ from the ICSD (Inorganic crystal structure database, ID: 4561, FIZ Karlsruhe) with the strongest intensity (101) and (102) peaks correspond to the $2\theta = 26.06°$ and $32.12°$. In our case, we have not seen any peak at these positions, confirming the absence of $B_2O_3$-related impurities. The FWHM of the (0002) peak reduces to ~0.303° (after the SPS) from ~0.503° (before the SPS), indicating improved crystallinity of h-BN.

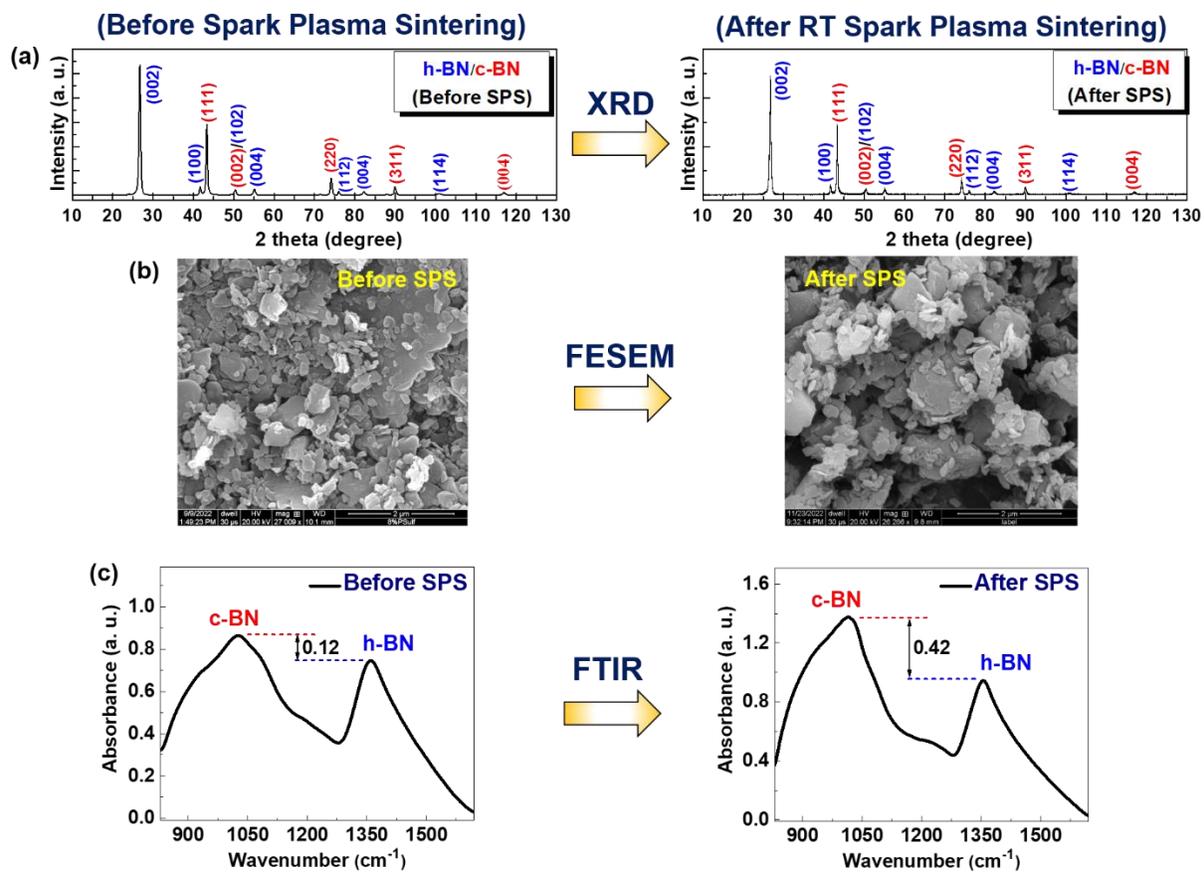

**Figure S5:** (a)-(c) XRD, FESEM, and FTIR of the h-BN/c-BN nanocomposite after the SPS done at room temperature (RT) and 90 MPa pressure.

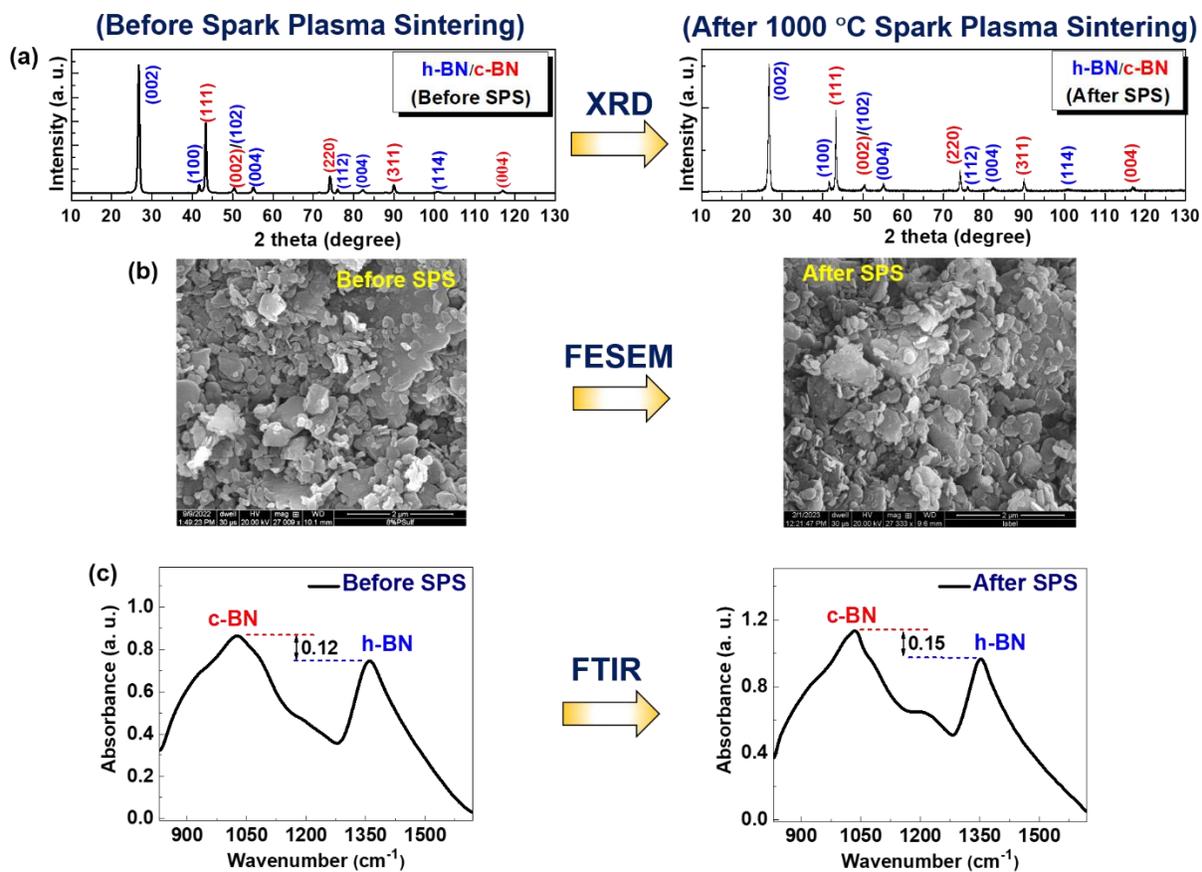

**Figure S6:** (a)-(c) XRD, FESEM, and FTIR of the h-BN/c-BN nanocomposite after the SPS done at 1000 °C and 90 MPa pressure.